\documentclass[aps,11pt,final,notitlepage,oneside,onecolumn,nobibnotes,
nofootinbib,preprintnumbers,superscriptaddress,showpacs,
centertags,showkeys,amsmath,amssymb]{revtex4}

\usepackage{epsfig}

\usepackage{graphicx} 

\usepackage{dcolumn} 

\def\phi{\varphi}

\begin{document}

\title{MONITORING OF CHARMED AND BEAUTY QUARK DISTRIBUTIONS IN PROTON AT LHC }

\author{G.I.Lykasov}

\altaffiliation[Also at ]{Institute for Nuclear Research, Dubna, Russia }

\email{lykasov@jinr.ru}

\author{V.A.Bednyakov}

\email{Vadim.Bednyakov@jinr.ru}

\affiliation{Joint Institute for Nuclear Research, 141980, Dubna, Russia}


\begin{abstract}
A short review on charmed and beauty hadron production in the lepton deep inelastic
scattering off proton, in proton-proton and proton-antiproton collisions 
at high energies is presented. It is shown that the existing theoretical and 
experimental information on charmed and beauty quark distributions 
in a proton is not satisfactory. A some procedure to study these
distributions at LHC energies is suggested. 

\end{abstract}



\maketitle

\section{Introduction}
Various approaches of the perturbative QCD including the next-to-leading order
calculations have been applied to construct the distributions of quarks in a
proton.  The theoretical analysis of the lepton deep inelastic scattering (DIS) off
protons and nuclei provide rather realistic information on distribution of
light quarks like $u,d,s$ in proton. However, to find a believable distribution of 
heavy quarks like $c({\bar c})$ and especially $b({\bar b})$ in proton
describing the experimental data on the DIS is non-trivial task. It is due
mainly to small values of $D$- and $B$- meson yields produced in the DIS at
existing energies. Even at the Tevatron energies the $B$- meson yield is not
so large. However, at LHC energies the multiplicity of $D$- and $B$- mesons
produced in $p-p$ collisions will be larger significantly. Therefore one can
try to extract a new information on the distribution of these heavy mesons in
proton. In this paper we suggest some proposal to study the distribution of
heavy quarks like $c({\bar c})$ and $b({\bar b})$ in proton based on the analysis
of the future LHC experimental data.

\section{\bf Hard parton scattering model}
Usually, the multiple hadron production in hadron-nucleon collisions at high
initial energies and large transfers is analyzed within the hard parton
scattering model (HPSM) suggested in Refs.\cite{FF,FFF} and Ref\cite{AVE:1974}. For example, 
the inclusive spectrum of hadron produced in the hard $p-p$ interaction is presented in the
following convolution form:
\begin{equation}
E_h\frac{d\sigma}{d^3p_h}=\sum_{q_f,g}\int G_{q_f,g}^{p_1}(x_1,k_{1t})G_{q_f,g}^{p_2}(x_2,k_{2t})
\frac{{\hat s}}{\pi}\frac{d\sigma({\hat s},{\hat t})}{d{\hat t}} 
\frac{1}{z^2}D_{q_f,g}^h(z)\delta({\hat s}+{\hat t}+{\hat u})dx d^2k_t dz~,
\label{def:hdsc}
\end{equation}
where $E_h,p_h$ are the energy and three-momentum of the produced hadron $h$;  
$G_{q_f,g}^{p_1}(x_1,k_{1t})$ is the distribution of quark with flavor
$f$ or gluon in the first colliding proton depending on the Feynman variable 
$x_1$ and the transverse momentum $p_{1t}$; $G_{q_f,g}^{p_2}(x_2,k_{2t})$ is 
the same quark or gluon distribution in the second colliding proton; ${\hat s},
{\hat t}$ and ${\hat u}$ are the Mandelstam variables corresponding to the
colliding quarks; $d\sigma({\hat s},{\hat t})/d{\hat t}$ is the 
differential cross section of the elastic parton-parton scattering; $z$ is 
the fraction of the produced hadron $h$ \cite{FF,FFF}.\\
The HPSM has been improved significantly applying
the QCD parton approach implemented in the modified minimal-subtraction
renormalization and factorization scheme. It provides a rigorous theoretical 
framework for a global data analysis. In this framework there are two distinct
approaches for next-to-leading order (NLO) calculations in perturbative QCD.

The first calculation scheme is the so-called massive scheme or fixed-flavor-number 
scheme (FFNS) developed in Refs.\cite{Nasson, Beenakker1} and Refs..\cite{Beenakker2,
Bojak}. In this approach the number of active flavors in the initial state is
limited to $n_f=4$, e.g., $u({\bar u}),d({\bar d}),s({\bar s})$ and $c({\bar c})$ 
quarks being the initial partons, whereas the $b({\bar b})$ quark appears only
in the final state. In this case the beauty quark is always treated as a
heavy particle, not as a parton. In this scheme the mass of heavy quarks acts
as cutoff parameter for the initial- and final-state collinear singularities
and sets the scale for perturbative calculations.   
Actually, the FFNS with $n_f=4$ is limited to a rather small range of transverse momenta 
$p_t$ of produced $D$ or $B$-mesons less than the masses of $c$ or $b$ quarks. In this 
scheme the $m_{c,b}^2/p_t^2$ terms are fully included.  

The another approach is the so-called zero-mass variable-flavor-number scheme
(ZM-VFNS), see Refs.\cite{Kniehl1,Kniehl2} and Ref.\cite{Greco} and references therein.
It is the conventional parton model approach, the zero-mass parton
approximation  is applied also to the $b$ quark, although its mass is
certainly much larger than the asymptotic scale parameter $\Lambda_{QCD}$.
In this approach the $b({\bar b})$  quark is treated as incoming parton
originating from colliding protons or proton-antiproton. This approach can be
used in the region of large transverse momenta of produced charmed or beauty
mesons, e.g., at $p_t\ge m_{c,b}$. Within this scheme the terms of order 
$m_{c,b}^2/p_t^2$ can be neglected. Recently the experimental inclusive $p_t$-
spectra of $B$-mesons in $p-{\bar p}$ collisions obtained by the CDF
Collaboration \cite{CDF1,CDF2} at the Tevatron energy $\sqrt{s}$=1.96 TeV 
in the rapidity region $-1\le y\le 1$ have been described rather satisfactory 
within this ZM-VFNS approach in Ref.\cite{Kniehl3} at $p_t\ge 10 (GeV/c)$
using the non-perturbative structure functions. In another kinematic region, e.g.,
at $2.5 (GeV/c\le p_t\le 10(GeV/c)$ the FFNS model mentioned above allowed to describe 
the CDF data without using fragmentation functions of $b$ quarks to $B$-mesons. 
Both these schemes have some uncertainties related to the renormalization parameters.
  
Looking the CDF data on the $p_t$-spectra of $B$-mesons produced in  $p-{\bar p}$ 
collision at the Tevatron energies published in Ref.\cite{CDF1} and
Ref.\cite{CDF2} one can find a difference between them especially at high values of  $p_t$.
Therefore it would be very useful to have more precise data at different LHC energies.

The experimental inclusive $p_t$-spectra of $D$-mesons obtained by
the CDF Collaboration at the same Tevatron energy and rapidity region \cite{CDF_Dmes}
at $p_t\ge 5$ (GeV/c) were described satisfactory within the so-called general-mass
variable-flavor-number scheme (GM-VFN) \cite{Kniehl4} not assuming the zero mass for
$c$ quarks. 

To calculate the inclusive spectrum of hadrons, for example heavy mesons exploring eq.(\ref{def:hdsc})
we have to know the distributions of quarks and gluons and their fragmentation functions (FF).
Usually, they are calculated within the QCD using the experimental information from
the DIS of leptons off protons and in the $e^+-e^-$ annihilation. The information on the gluon distribution
and its fragmentation function can be extracted from the experimental data on the jet production \cite{CDF_jet} 
and their theoretical analysis \cite{Giele:1993,Pumplin:2002}. Unfortunately the theoretical QCD calculation
of the jet production has different sources of uncertainty. As is shown in Ref.\cite{CDF_jet} the main 
contribution comes from uncertainty on parton distribution functions (PDFs) and is computed within the 
method suggested in Ref.\cite{Pumplin2:2002}. At low transverse momenta of jets $p_t^{jet}$ the uncertainty
is small and approximately independent on the jet rapidity $y^{jet}$. The uncertainty increases as $p_t^{jet}$
and  $\mid y^{jet}\mid$ increase. It can become about $130 $ percents \cite{CDF_jet}. To analyze the 
jet- and heavy quark-production at low $p_t$ and large $y$ or large values of Feynman 
variable $x_F$ one can apply the another nonperturbing QCD model, the so-called Quark-Gluon String Model.  
(QGSM).

\section{\bf The Quark-Gluon String Model}
The QGSM is based on the $1/N$ expansion in QCD suggested in Refs.\cite{tHooft,Veneziano} 
instead of $\alpha_s$ expansion that has the infrared divergence problem at $Q^2\rightarrow 0§$,
here $N$ is the number of flavors or colors. The relation of the topological expansion over
$1/N$ of the hadron-hadron scattering amplitude to its $t$-channel one over Regge poles has been
suggested in Refs.\cite{Kogut,Marchesini}. This approach has been applied
to analyze soft hadron processes at high energies, see for example Ref.\cite{Kaid1}.  

It has been shown \cite{Kaid1} that the main contribution to the inclusive spectrum of hadron produced in
$p-p$ collisions at high energies comes from the so-called cylinder graphs corresponding to one-Pomeron 
and multi-Pomeron exchanges which are presented in Fig.1.
 
\begin{figure}[htb]
{\epsfig{file=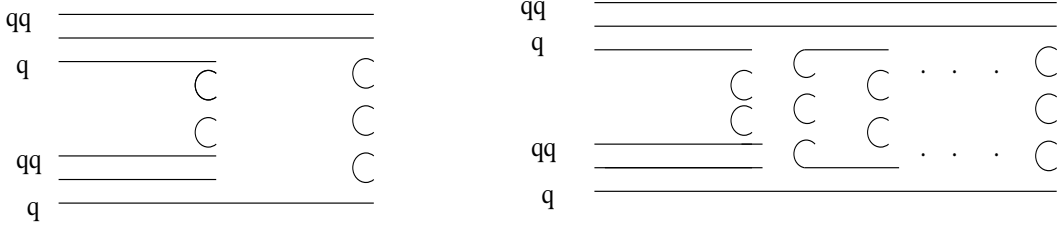,height=3.cm,width=14.cm }}
\caption[Fig.1]{The cylinder graph corresponding to the one-Pomeron
exchange (left diagram) and the multi-cylinder
graph corresponding to the multi-Pomeron exchanges (right diagram).} 
\end{figure}
According to the QGSM,  between quark $q$
(diquark $qq$) and diquark $qq$ (quark $q$) for colliding protons (Fig.1, left
graph) the colorless strings are formed, then after their brake
$q{\bar q}$ pairs are created which fragmentate to the hadron
$h$. The right diagram in Fig.1 corresponds to the creation of two colorless 
quark-diquark and diquark-quark strings and $2(n-1)$ chains between see quark (antiquark)
and antiquark(quark). 
The inclusive spectrum $\rho_{h}$ of hadron $h$
produced in $p-p$ collision corresponding to the one-Pomeron graph
(Fig.1, left graph) has the following form \cite{LAS,LS:1996}:
\begin{equation}
\rho_{h}(x,{\bf p}_t)=\sigma_1 F_q^{h}(x_+,p_t) F_{qq}^{h}(x_-,{\bf p}_t)/F_{qq}^{h}(0,{\bf p}_t)+
F_{qq}^{h}(x_+,{\bf p}_t)F_q^{h}(x_-,p_t)/F_q^{h}(0,p_t)~,
\end{equation}
where $\sigma_1$ is the cross section of the 2-chains production, corresponding
to the $s$-channel discontinuity of the cylinder (one-Pomeron) graph, usually it
is calculated within the quasi-eikonal approximation \cite{Termart};
\begin{equation} 
F_{q(qq)}^{h}(x_{\pm},{\bf p}_t)=\sum_{flavors}\int_{x_{\pm}}^1 dx_1\int
d^2p_{1t} \ d^2p_{2t} \ f_{q(qq)}(x_1,{\bf p}_{1t}) \
G_{q(qq)}^{h}(\frac{x_\pm}{x_1},{\bf p}_{2t}) 
\delta^{(2)}({\bf p}_{1t}+{\bf p}_{2t}-{\bf p}_t)~,
\label{def:rhoh} 
\end{equation}
Here $f_{q(qq)}(x_1,{\bf p}_{1t})$ is the quark or diquark distribution function
depending on the Feynman variable $x_1$ and the transverse momentum of quark or diquark,
$z^{-1}G_{q(qq)}^{h}(z,{\bf p}_{2t})=D_{q(qq)}^{h}(z,{\bf p}_{2t})$ 
is the fragmentation function of  quark or diquark to the hadron $h$;
$x_\pm=\frac{1}{2}(\sqrt{x_t^2+x^2}\pm x)$ and $x_t=2m_{h t}
/\sqrt{s},~m_{h t}= \sqrt{m_{h}^2+p_{t}^2}$,
$s$ is the square of the initial energy in the $p-p$ c.m.s. Actually,
the interaction function $F_{q(qq)}^{h}(x_+,p_t)$ corresponds to the
fragmentation of the upper quark (diquark) to the hadron $h$,
whereas $F_{qq(q)}^{h}(x_-,p_t)$ corresponds to the fragmentation of
the down diquark (quark) to $h$, see Fig.1 (left diagram). 
The expression for the inclusive spectrum of the hadron $h$
produced in $p-p$ collision corresponding to the right graph of Fig.2
has more complicated form, see the details in Refs.\cite{LAS,LS:1996}.

All the quark distributions and fragmentation functions are related to the intercepts and slopes
of Regge trajectories. For example,the distribution of $c({\bar c})$ quarks in a proton obtained 
within the QGSM and the Regge theory has the following form \cite{LAS}:
\begin{equation}
f_{c({\bar c})}^{(n)}=C_{see}^{(n)}\delta_{c({\bar c})}x^{-\alpha_\psi(0)}
(1-x)^{\alpha_R(0)-2\alpha_N(0)+(\alpha_R(0)-\alpha_\psi(0)+n-1}~,
\end{equation}
where $n$ is the number of Pomeron exchanges, $\alpha_R(0),\alpha_N(0)$ are the intercepts of the
Reggeon and nucleon Regge trajectories, $\alpha_\psi(0)$ is the intercept of the $\psi$-Regge trajectory,  
$\delta_{c({\bar c})}$ is a probability fraction 
of $c({\bar c})$ pairs in a quark see of the proton.\\ 
The intercepts $\alpha_R(0)$ and $\alpha_N(0)$ are known very well from the experimental data on
the soft hadron processes and the corresponding Regge trajectories have linear behavior as a 
function of the transfer $t$. As for the Regge trajectory $\alpha_\psi(t)$, the information on its 
$t$-dependence is rather poor. As a function of $t$, it can be linear or nonlinear. 

The $b({\bar b})$ quark distributions and the fragmentation functions of $b({\bar b})$ 
to $B$-mesons can be also obtained within the Regge theory and the QGSM. They are related to the conventional 
Regge trajectories of light mesons and the Regge trajectory of the $\Upsilon$-meson consisting of $b{\bar b}$ pair. 
The information on the $t$-dependence of the $\Upsilon$-Regge trajectory is also uncertain. 

The modified version of the QGSM suggested in Refs.\cite{LAS,LS:1996}
including the transverse momentum dependence of the interaction functions $F_{q,qq}^h(x_\pm,p_t)$
allowed to describe the experimental data on inclusive spectra of $D$-mesons produced in $p-p$ collisions 
at the ISR energies and at moderate values of transverse momenta until $p_t\simeq 4-5 (GeV/c)$. 
However, the results are very sensitive to the value of the intercept $\alpha_\psi(0)$.
Note, that the QGSM in contrast to the perturbative QCD has not uncertainty related to the mass parameter.

\section{\bf PROPOSAL}
Concluding a short review on the HPMS and the QGSM we would like to propose 
to do a complex theoretical analysis both the jet production in $p-p$ and $p-{\bar p}$ collisions 
within the perturbative QCD and the semi-hard production of $D$- and $B$-mesons in these reactions within the 
modified version of the QGSM. The semi-hard hadron processes mean the inclusive production of heavy mesons at 
not large transverse momenta $p_t$ and not small values of $x_F$. 
We are going to extract the gluon distribution in a proton and its fragmentation
functions to heavy mesons from the QCD analysis of the jet production in $p-p$ and $p-{\bar p}$ inelastic
processes. Constructing a new modified version of the QGSM based on Refs.\cite{LAS,LS:1996} we intent to 
include also the gluon-gluon and gluon-quark interactions using the obtained gluon distributions in proton and 
its FF. Then we will apply the suggested approach to describe all the existing experimental data on $D$- and 
$B$-mesons produced in $p-p$ and $p-{\bar p}$ collisions at high energies and perform corresponding predictions at 
LHC energies. From this analysis we propose to extract a new information on the distribution of charmed and beauty 
quarks in proton.

\begin{acknowledgments}
We are grateful to A.B.Kaidalov, A.V.Efremov and V.Kim for very useful discussions. 
V.B. acknowledges the support of the RFBR (grant RFBR-06-02-04003). 

\end{acknowledgments}

\end{document}